# Research On Permanent Magnet BLDC for small electric vehicle

Uma Gupta
Dept. of Electrical and Electronic engineering, Amity University
sec-125, Noida, Uttar Pradesh.
umagupta14@gmail.com

**Abstract-** In this paper, different electric motors are studied and compared to see the benefits of each motor and the one that is more suitable to be used in the electric vehicle (EV) applications. There are five main electric motor types, DC, induction, permanent magnet synchronous, switched reluctance and brushless DC motors are studied. It is concluded that although the induction motors technology is more mature than others, for the EV applications the brushless DC and permanent magnet motors are more suitable than others. The use of these motors will result in less portion, less fuel consumption and higher power to volume ratio. The reducing prices of the permanent magnet materials and the trend of increasing efficiency in the permanent magnet and brushless DC motors make them more and more attractive for the EV applications. This paper first describes the main features of BLDC control methods, then after comparison it's parameters will be controlled using Field Oriented Control or Vector control method.

## INTRODUCTION

Permanent Magnet (PM) motors are nowadays regarded as an interesting solution for variable-speed drives which are faded with all ranges of inverters. The interest in these motors has been increased because of their feasibility for vehicle propulsion. Consequently, they are also of good for small electric drives applications. The most important advantages that are expected in comparison to the asynchronous motors are lower losses and a higher torque density. The thesis presents the study and the design analysis of a permanent magnet brushless dc motor for small vehicle drive. Various aspects of the design of PM motor drives are under consideration with special attention to the requirement of maximum torque and field weakening concepts among all these aspects particularly, the sensor-less control or vector control without a mechanical rotor position sensor, is considered in this paper. The term sensorless does not represent the lack of sensors entirely, but the fact that in comparison with other drives from the same category of FOC, it denotes that the speed and/or position sensor is missing. This feature decreases the cost of the drive system which is always desired, but this is not the only reason for this approach, as some applications have requirements concerning the size and lack of additional wiring for sensors or devices mounted on the shaft (due to hostile environments such as high temperature, corrosive contacts e.t.c.).So the field oriented control (FOC) technique of sensor less control is used in this paper. An advantage of FOC is that it is a less complex algorithm and as it is a vector control technique so angle of the rotor can also be identified along with the magnitude. As in case of motor the frequency, amplitude, toque and angle of rotation matters so all these purposes can be easily solved with the use of FOC. The scope of the paper is to simulate the FOC.

The so called brushless DC (BLDC) motors are broadly used as actuators. PERMANENT-MAGNET (PM) motors have been the choices for electrical vehicle (EV) applications due to their high efficiency, compact size, high torque at low speeds, and ease of control regenerative breaking [5]. Since the permanent magnet synchronous motor has the feature of small size, high efficiency and superior reliability, it is widely used in many important areas such as modern industrial automation, military, chemical industry, aviation and aerospace [1-4]. PM motors with higher power densities are also now increasingly choices for aircraft, marine, naval, and space applications. Permanent magnet brushless motors for electric vehicles are presently given an increasing attention. With this trend, this paper deals with an interior permanent magnet (IPM) brushless DC motor (BLDCM).

In this work, the simulation of a field oriented controlled Permanent Magnet Brushless DC (PMBLDC) motor drive system is developed using Simplorer. For simplicity purpose the Simplorer software is used in this thesis. The simulation circuit will include all realistic components of the drive system. This enables the calculation of currents and voltages in different parts of the inverter and motor under transient and steady state conditions. A closed loop control system with a PI controller in the speed loop has been designed to operate in constant torque and flux weakening regions. Implementation has been done in Simplorer. A comparative study of hysteresis and PWM control schemes associated with Space Vector Modulation (SVM) has been made in terms of Harmonic spectrum and to the harmonic distortion. Simulation results are given for two speed of operation, one below rated and another above rated speed.

## Control Techniques

A wide range of control algorithms are available:

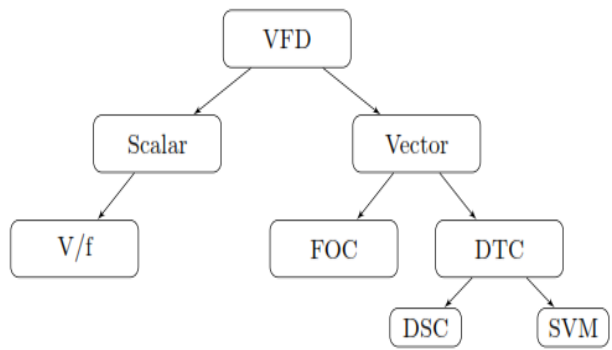

Figure A. Control Techniques

1. Trapezoidal control:
 Also known as six-step control, this is the simplest algorithm. For each of the six commutation steps, a current path is formed between a pair of windings, leaving the third winding disconnected. This method generates high torque ripple, leading to the vibration, noise, and poorer performance compared to other algorithms.
2. Sinusoidal control:
Also known as voltage-over-frequency commutation, sinusoidal control it overcomes many of the issues involved with trapezoidal control by supplying smoothly (sinusoidal) varying current to the 3 windings, thus reducing the torque ripple and offering a smooth rotation. However, these time-varying currents are controlled using basic PI regulators, which lead to poor performance at higher speeds.
3. Field Oriented Control (FOC):
Also known as vector control, FOC provides better efficiency at higher speeds than sinusoidal control. It also guarantees optimized efficiency even during transient operation by perfectly maintaining the stator and rotor fluxes. FOC also gives better performance on dynamic load changes when compared to all other techniques.
4. Scalar control:
 Scalar control (or V/Hz control) is a simple technique to control speed of induction motor.
In this paper Vector Control or Field Oriented Control (FOC) is used because of it's following advantages-

- Transformation of a complex and coupled AC model into a simple linear system
- Independent control of torque and flux, similar to a DC motor
- Fast dynamic response and good transient and steady state performance
- High torque and low current at start-up
- High Efficiency
- Wide speed range through field weakening

## Vector Control of Motor

At the core of vector control algorithm are two important mathematical transforms: Clark Transform, Park Transforms and their inverse. Use of the Clark and Park transforms bring the stator currents which can be controlled into the rotor domain. Doing this allows a motor control system to determine the voltages that should be supplied to the stator to maximize the torque under dynamically changing loads.

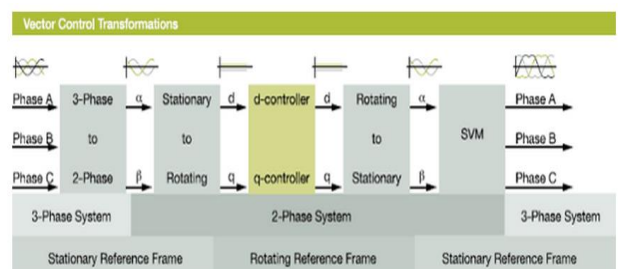

Figure B. Vector Control Transformation

1. Clark Transformation:
The Clark mathematical transform modifies a three phase system to a two co-ordinate system.

$$i_\alpha = \frac{2}{3} \cdot i_a - \frac{1}{3}(i_b - i_c)$$
$$i_\beta = \frac{2}{\sqrt{3}}(i_b - i_c)$$
$$i_o = \frac{2}{3}(i_a + i_b + i_c)$$

Where Iα, Iβ are components of the orthogonal reference plane and I₀ is the homo planer component which is of little significance.

2. Park Transformation:

The Park mathematical transform converts two-phase stationary system vectors to rotating system vectors

$$i_{sd} = i_\alpha \cdot \cos(\theta) + i_\beta \cdot \sin(\theta)$$
$$i_{sq} = -i_\alpha \cdot \sin(\theta) + i_\beta \cdot \cos(\theta)$$

The two phases α, β frame representation calculated with the Clarke transform is then fed to a vector rotation block where it is rotated over an angle θ to follow the d, q frame attached to the rotor flux. The rotation over an angle θ is done according to the above formulas.

These two transformations are shown in figure B.

## Basic Scheme for Vector Control of AC Motor

Figure C. shows the basic scheme for a Field Oriented Vector Control for an AC Motor. The Clarke transform uses three-phase currents IA, IB and IC to calculate currents in the two-phase orthogonal stator axis: Iα and Iβ. These two currents in the fixed coordinate stator phase are transformed to the Isd and Isq currents components in the d, q frame with the Park transform. The currents Isd, Isq and the instantaneous flux angle θ, calculated by the motor flux model, are used to calculate the electric torque of an AC motor.

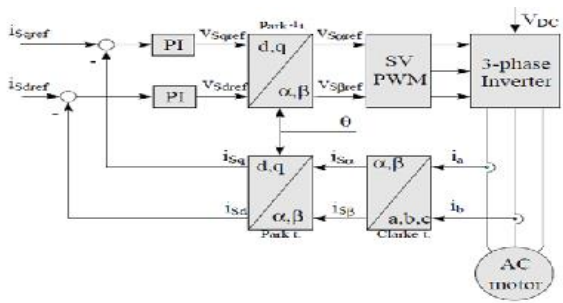

Figure C. Basic Scheme for Vector Control of AC Motors

## Space Vector Modulation (SVM) Invertor

Voltage source inverters are devices that convert a DC voltage to AC voltage of variable frequency and magnitude. They are commonly used in adjustable speed drives and are characterized by a well-defined switched voltage waveform in the terminals [6]. Figure D shows a voltage source inverter. The AC voltage frequency can be constant or variable.

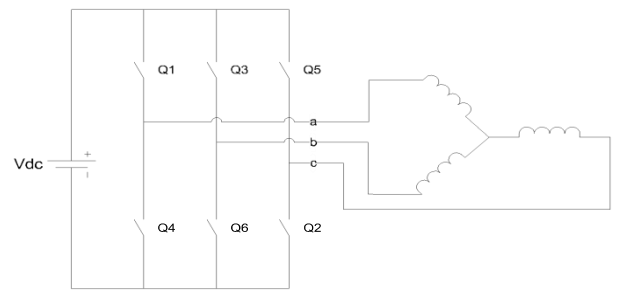

Figure D. Voltage Source Inverter Connected to a Motor

Three phase inverters consist of six power switches connected as shown in figure D to a DC voltage source. The inverter switches must be carefully selected based on the requirements of operation, ratings and the application.

Space vector modulation working Principle

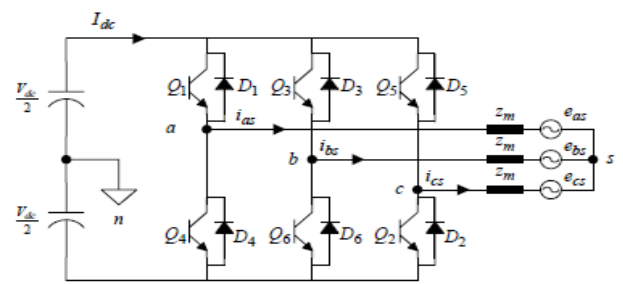

Figure E. Topology of a basic three phase inverter

A three phase inverter as shown in fig. E converts the DC supply, via a series of the switches to three output legs which could be connected to a three-phase motor.

The switches must be controlled so that at no time are both switches in the same leg turned on or else the DC supply would be shorted. This requirement may be met by the complementary operation of the switches within a leg. i.e. if A+ is on then A− is off and vice versa. This leads to eight possible switching vectors for the inverter, V0 through V7 with six active switching vectors and two zero vectors. This Mechanism can be shown as fig.F

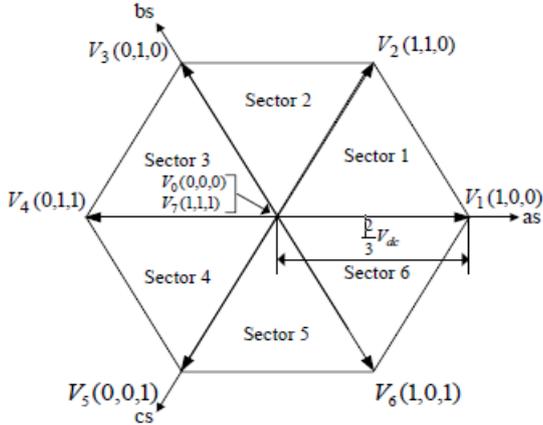

Figure F. Six vector representation

**Field-Weakening Operation In Motors**

Motor characteristics of a separately excited DC commutator (or a synchronous motor) can be used to explain the field-weakening operation. The excitation flux is controlled by varying the dc current through the field winding. The torque is the product of the armature current and the excitation flux in the d-axis. The induced voltage is proportional to the speed and flux. At low speed, the rated current I and the rated flux Φ are used to generate the rated torque. The voltage and the output power both increase linearly with the speed. A rated speed is reached when the voltage equals to the rated voltage that is defined as the maximum voltage available from the drive. This rated speed is often referred to as the base speed. Therefore, in order to increase the speed above the rated speed, the flux must be decreased or weakened whilst the voltage is kept constant at rated value. The torque is inversely proportional to the speed so that the output power remains constant beyond the base speed. As shown in figure D., this operating region is called the flux-weakening or field-weakening region.

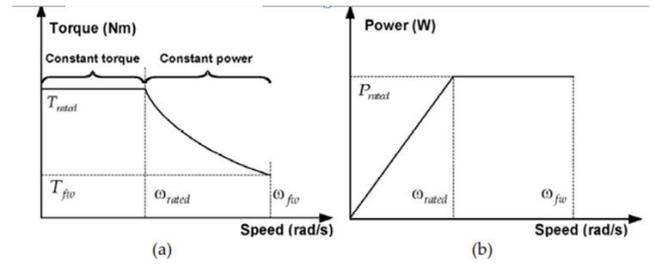

Figure G. Ideal field weakening characteristics :
(a)Torque characteristic (b) Power characteristic

Field-weakening Operation of Permanent Magnet Motors:

In permanent magnet motors, the excitation flux is produced by the magnets. As the magnets resemble a "fixed excitation flux" source, the magnetic field cannot be varied as in a separately excited DC motor by controlling the field current. However, a control of the total flux in the d-axis (or field-weakening) is achieved by introducing an armature flux against the fixed excitation field from the magnets. It is achieved by injecting a negative d-axis current Id, as illustrated in fig.H. This can be further elucidated with simple vector diagrams of a non-salient PM motor. When the motor is operating at the rated condition with the maximum possible voltage Vb, it can be noted that the voltage vector is on the voltage limit contour. It is virtually impossible to increase the speed further once the voltage limit is reached. In order to increase the speed beyond the rated speed, an introduction of the negative d-axis current Id is then necessary. Wsith the help of the imposing current Id, the voltage vector V is "brought back" within the voltage limit. The magnitude of Id is gradually increased as the current angle γ varies from 0 to π/2 electrical radians. The voltage limit Vb of the PM motor can be expressed as

$$V_b^2 \geq \omega^2\left[\left(\Psi_m + L_d I_d\right)^2 + \left(L_q I_q\right)^2\right]$$

where ω is the electrical operating speed, Ψm is the magnet flux, Ld and Lqare the d-axis and q-axis inductance respectively. The saliency ratio is defined as the ratio between the q-axis inductance (Lq) over the d-axis inductance (Ld). Depending on the rotor configuration, a PM motor is then referred to salient when Lq is not equal to Ld, and non-salient if Lqis equal to Ld. Saliency ratio plays a significant role on the field-weakening performance as it will be shown in the subsequent reviewed literatures.

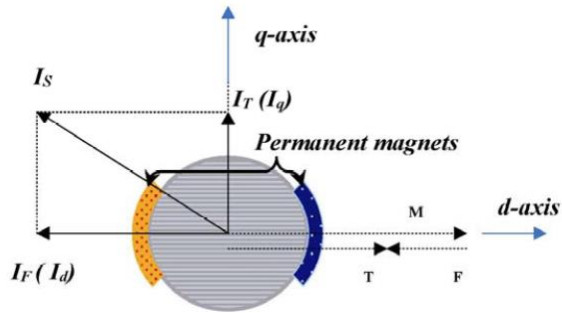

Figure H. Flux-weakening of permanent magnet motors.

## Simplorer Results Of Interior PM BLDC Motor

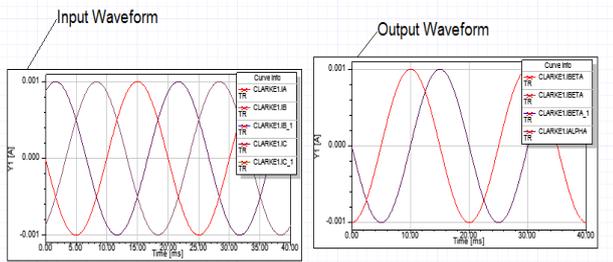

Figure I. Input-Output waveform of Clarke Transformation

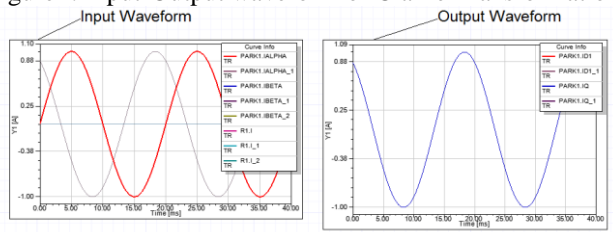

Figure J. Input-Output waveform of park Transformation

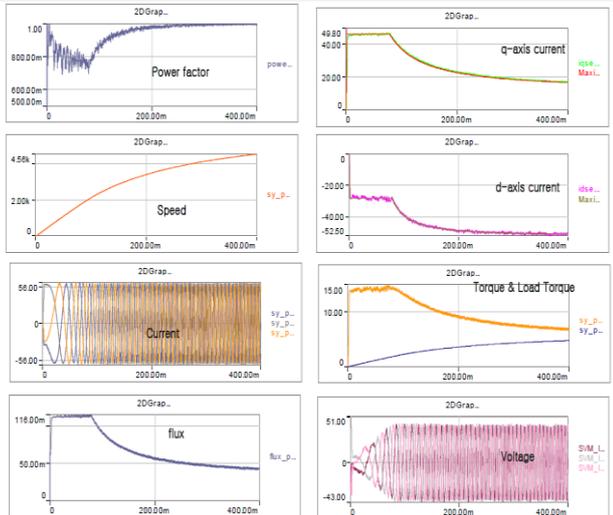

Figure K. Power factor, Speed, Torque, Current, Flux Voltage waveforms

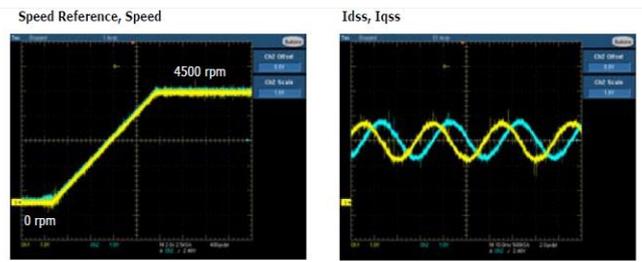

Figure L. Speed output and Current output at Stationary reference frame

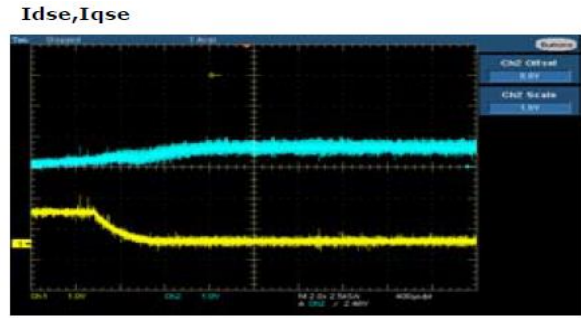

Figure M. Current outputs for d, q planes at Rotary reference frame

## CONCLUSION

In this research work the performance of interior Permanent Magnet Brushless DC is studied theoretically as well practically. With the use of SIMPLORER software a detailed Simulink model has being developed and operation below and above rated speed has been studied. Simplorer has been chosen from several simulation tools because its flexibility in working with analog and digital devices. Mathematical models can be easily incorporated in the simulation and the presence of numerous tool boxes and support guides simplifies the simulation of large system compared to Spice. Simplorer is capable to show in real time results with reduced simulation time and debugging.

In the present simulation measurement of currents and voltages in each part of the system is possible, thus permitting the calculation of instantaneous or average losses, efficiency of the drive system and total harmonic distortion.

The type of motor control technique used here was- vector control technique or field oriented control. A faster response has been seen of FOC for BLDC. It can be concluded that FOC is better other than other due to following points-

1.Only FOC is capable to control torque and flux at very low speed.

2.In v/hz and scalar techniques the current increase rapidly which makes the motor less durable.

3. The noise level is high at low speed in the control techniques other than FOC.
4. Variable switching frequency behavior is not found in FOC.
5. In the control techniques other than FOC, the DC current control is very less.

So the aim of parameters control of PM BLDC motor has been found successfully with FOC at SIMPLORER software.